\documentclass[
12pt,
3p,
review,
]{elsarticle}
\usepackage{hyperref}
\usepackage{graphicx,graphics}
\usepackage{placeins}
\usepackage{amsmath}
\usepackage{amssymb}
\usepackage{mathtools}
\usepackage{array}
\usepackage{hyperref}
\usepackage{color}
\usepackage{soul}
\usepackage{dcolumn}
\usepackage{bm}
\usepackage{todonotes}
\usepackage[utf8]{inputenc}
\usepackage[english]{babel}
\usepackage{libertine}
\usepackage{pgfplots}
\usepgfplotslibrary{polar}
\pgfplotsset{compat=1.9}
\usetikzlibrary{calc}

\usepackage[title]{appendix}

\usepackage{multirow}

\usepackage{layouts}

\usepackage{caption}
\usepackage{subcaption}

\renewcommand{\l}{\mathopen{}\mathclose\bgroup\left}
\renewcommand{\r}{\aftergroup\egroup\right}

\let\originaleps=\epsilon
\let\epsilon=\varepsilon
\let\varepsilon=\originaleps
\usepackage{stmaryrd}

\mathchardef\hy="2D

\definecolor{mydark_blue}{RGB}{0, 0, 139}
\definecolor{myblue}{RGB}{0, 0, 255}
\definecolor{mycyan}{RGB}{0, 255, 255}  
\definecolor{mygreen}{RGB}{0, 255, 0}
\definecolor{myyellow}{RGB}{255, 255, 0}
\definecolor{myred}{RGB}{255, 0, 0}
\definecolor{mydark_red}{RGB}{139, 0, 0}
\definecolor{myblack}{RGB}{0, 0, 0}

\definecolor{BRY_1}{RGB}{  0,  0,255}
\definecolor{BRY_2}{RGB}{127,  0,127}
\definecolor{BRY_3}{RGB}{255,  0,  0}
\definecolor{BRY_4}{RGB}{255,127,  0}
\definecolor{BRY_5}{RGB}{255,255, 85}

\journal{Materialia}

\begin{document}

\begin{frontmatter}
\title{Accessing topological feature of polycrystalline microstructure using object detection technique}

\author[mymainaddress]{Mridhula Venkatanarayanan}
\author[mymainaddress,mysecondaryaddress]{P G Kubendran Amos\corref{mycorrespondingauthor}}
\ead{prince@nitt.edu}

\cortext[mycorrespondingauthor]{P G Kubendran Amos}

\address[mymainaddress]{Theoretical Metallurgy Group,
Department of Metallurgical and Materials Engineering,\\
National Institute of Technology Tiruchirappalli, \\
Tamil Nadu, India}

\address[mysecondaryaddress]{Institute of Applied Materials (IAM-MMS),
Karlsruhe Institute of Technology (KIT),\\
Strasse am Forum 7, 76131 Karlsruhe, Germany
}

\begin{abstract} 

Faces-classes of grains, often referred to as topological features, largely dictate the evolution of polycrystalline microstructures during grain growth. 
Realising these topological features is generally an arduous task, often demanding sophisticated techniques. 
In the present work, a distinct machine-learning algorithm is extended for the first time to comprehend the topological distribution of the grains constituting a polycrystalline continuum. 
This regression-based object-detection approach, besides significantly reducing human-efforts and ensuring computational efficiency, predicts the face-class of the grains by introducing appropriate \lq bounding boxes\rq \thinspace.
After sufficient training and validation, over $500$ epochs, the current model exhibits a remarkable overlap with the ground truth that encompasses manually realised topological features of the polycrystalline microstructures. 
Accuracy of this treatment is further substantiated by relevant statistical studies including precision-recall analysis.
The model is exposed to unknown test dataset and its performance is assessed by comparing its predictions with the labelled microstructures.
Reflecting the statistical accuracy, a strong agreement between the algorithm-predictions and the ground truth is noticeable in these comparative studies involving polycrystalline systems with varying number of grains.

\end{abstract}


\end{frontmatter}


\section{Introduction}

Besides phase transformations, microstructural changes include morphological evolution of the constituent phases driven by the ability of the system to reduce the interfacial-energy density. 
The overtly energy-minimisation changes, that are not accompanied by any change in the existing phase fraction, manifest in varied form depending on the characteristic feature of the microstructure. 
In system with distinguishable matrix and precipitate, coarsening reduces the interfacial energy per unit volume~\cite{deschamps2021precipitation}.
On the other hand, in polycrystalline systems, the interfacial-energy density is reduced by the overall increase in the average size of the grains rendered by the growth of larger grains at the expense of smaller ones~\cite{rohrer2005influence}. 
Despite the rather straightforward, the effect of energy-minising microstructural changes on the properties of the materials are noticeable~\cite{armstrong1970influence}. 
Consequently, numerous attempts have been made to understand and predict the kinetics of these transformations, particularly in polycrystalline systems~\cite{kim2006computer,perumal2020quadrijunctions,amos2020multiphase}. 

The energy-minising transformation, in polycrystalline systems, is induced by the curvature-difference in the inherent morphology of the grains. 
Correspondingly, the kinetics of grain growth is dictated primarily by the integral of the mean curvature over the walls that contribute to the morphology of the grains~\cite{mullins1956two,glazier1993grain}. 
In other words, growth kinetics are essentially governed by the walls of the grains and vary significantly with the change in the number of faces. 
The seminal expression describing the kinetics of grain growth in two-dimensional isotropic systems, incidentally, replaces the integral of mean curvature with number of faces~\cite{von1952metal}.
The extension of this expression to three-dimensional isotropic polycrystalline microstructures, though includes size of the grains, the role of face-class, \textit{id est} number of faces, continues to remain dominant~\cite{macpherson2007neumann}.

In addition to dictating the growth rate, the face class of a grain directly indicates its surrounding. 
Stated otherwise, the continuum in a microstructure is established by the grains sharing their sides (faces) with immediate neighbours. 
Accordingly, a face class further represents the number of neighbours encapsulating a given grain. 
The local evolution of a grain, beside being controlled by the curvature of the faces, it also effected by its neighbour~\cite{patterson2022relationship}. 
A face class then, in addition to offering a direct influence, is indicative of the role of neighbouring grains in the temporal change~\cite{wakai2000three}.

Treating the grains in a polycrystalline system analogously to space filling cellular structures, statically relates the size to the number of sides~\cite{rivier1985statistical}. 
Such insight relating the size and face class, which are contextually referred to \textit{geometrical} and \textit{topological} features, lends itself in quantitatively delineating the distribution of grains in a polycrystalline microstructure.
To that end, one primary criterion for re-creating realistic microstructure in simulation setup involves establishing the appropriate distribution of face classes in the system~\cite{ullah2017simulations}.
The topological features, therefore, characterise a polycrystalline system both locally and globally, through the immediate neighbours and the number-density distribution of the various face classes, respectively. 

Given the extended importance of the topological features in dictating the growth kinetics and grain distribution in polycrystalline microstructure, numerous techniques have been adopted to comprehend the face classes of the grains. 
Sophisticated sequence of treatments are developed to recreate the topology of the grains in a three-dimensional polycrystalline system~\cite{alkemper2001quantitative,groeber20063d,bhandari20073d,ullah2013optimal}.
With the advent of artificial intelligence and other related techniques, human efforts in deciphering complex images have significantly been reduced~\cite{sha2020artificial,amos2022high}. 
Consequently, these approaches are recently adopted to understand and classify microstructures~\cite{decost2015computer,azimi2018advanced}. 
Despite the progressive involvement of the deep learning treatments in microstructural analyses, their contribution in investigating polycrystalline systems has thus far been marginal.
Therefore, in the present work, regression-based object counting technique is extended to quantify the topological features of the grains in the polycrystalline system.

\section{Topology-based realisation of grains}

The object-detection has thus far been largely adopted in \lq non-technical\rq \thinspace, yet equally complex, applications~\cite{zou2019object}. 
An extension of this treatment to realise grains in polycrystalline systems, primarily based on their topological features, begins with a parallel perception of the microstructure that is relatable to the underlying algorithm.
Correspondingly, the grains constituting a polycrystalline microstructure are viewed as individual objects analogous to the foreground.
However, as opposed to a conventional image, wherein generally the background is noticeably different, in polycrystalline microstructure, the background surrounding the focused object continue to remain other similar objects, \textit{id est} grains.
The performance of object-detection treatment in systems exclusively comprising of similar objects, distinguished only by their features, have not been reported yet.  
Therefore, besides offering a technique for realising the topological features of the grains, the present work, by exposing a well-known object-detection technique, YOLOv5, to polycrystalline microstructure attempts to assess and expand its applicability~\cite{thuan2021evolution}. 
Furthermore, the established distinction of grains based on their number of sides incidentally translate to \textit{object-class} in the current analysis.

\subsection{Extending objection-detection treatment}

\begin{figure}
    \centering
      \begin{tabular}{@{}c@{}}
      \includegraphics[width=0.9\textwidth]{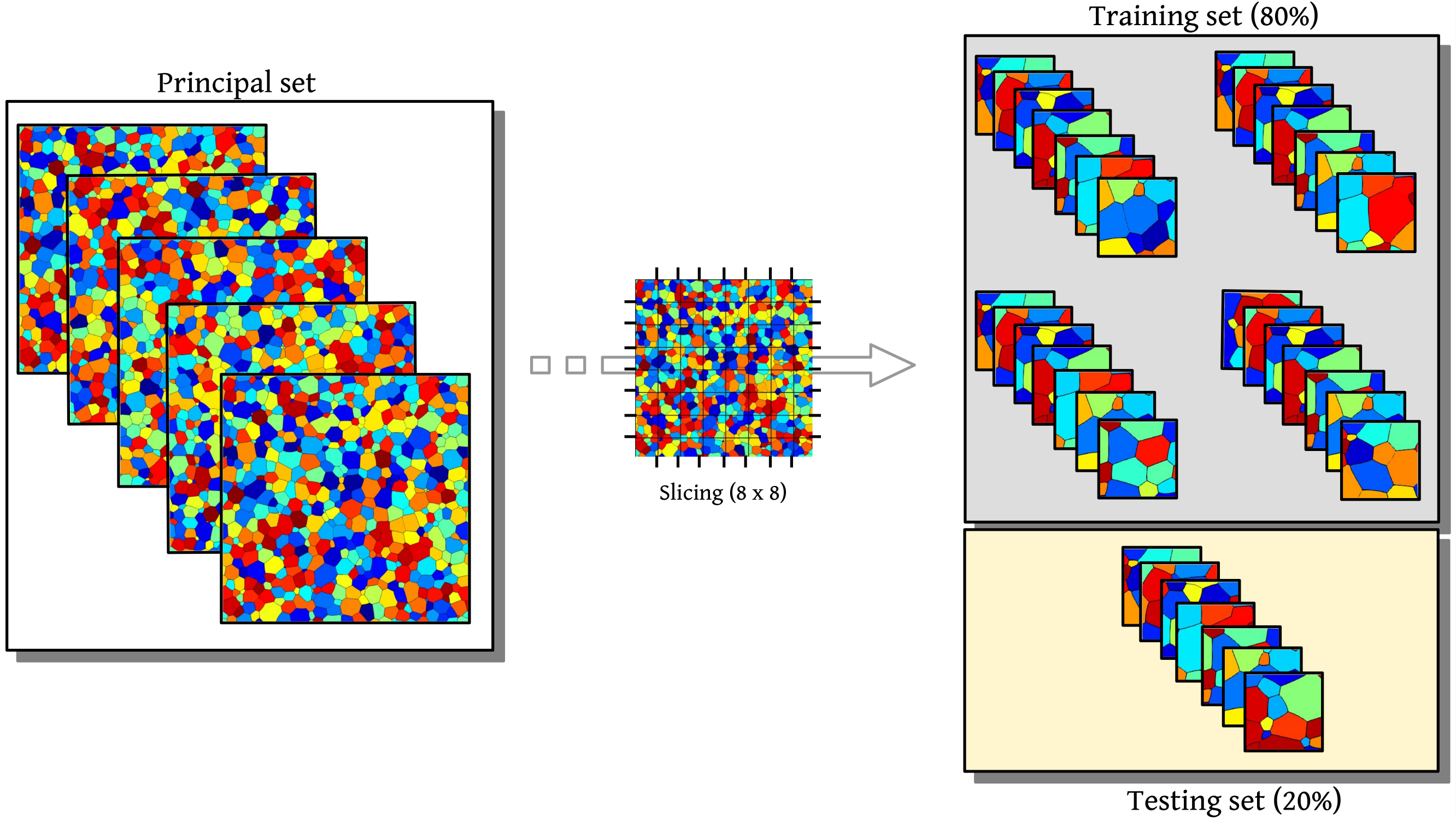}
    \end{tabular}
    \caption{  Schematic representation of the data acquisition scheme employed for generating training and test dataset. Besides numerically generated images, experimentally-observed microstructures gathered from open online searches have been included in training and test dataset (presented in supplementary document).
    \label{fig:fig1}}
\end{figure}

Though apparently different from a conventional machine-learning based data-analysis, the development of the present model adheres to the framework of a \textit{supervised} treatment. 
However, the key difference is that, in the current treatment, polycrystalline microstructures replace conventional numerical and/or categorical data, and are subsequently analysed. 
Its adherence to supervised learning means that the accuracy of the present topology-detection technique is principally dictated by the wealth of the dataset.
The required information to develop the technique is rendered as numerically-generated polycrystalline microstructure. 
Despite its largely theoretical conception, the proximity of the generated polycrystalline microstructures to the physical systems is ensured by adopting the proven Poisson-Voronoi tessellation~\cite{kumar1992properties}, expressed as
\begin{equation}
 R_i=\{\textbf{x}\in D ~ : ~ ||\textbf{P}_i -  \textbf{x}|| < ||\textbf{P}_j -  \textbf{x}||\quad \forall i\neq j \quad i,j=1,2,\dots,n \}.
\end{equation}
Under the above tessellation scheme, the domain $D\in \mathbb{R}^d$, where $d=2$ all through the present study, is partitioned into distinct regions $R_i$ through the seed points $\textbf{P}_i$. 
These seed points are randomly placed in the domains and are allowed to grow into regions (grains) till they impinge on a neighbour. 
The section wherein the two regions intersect form the \textit{face}, with the \textit{triple junctions} formed by the meeting of three regions. 
After establishing a continuum, the regions essentially appear as the grains in a polycrystalline system. 
In order to further ensure the morphology of these grains are indicative of physical systems, and are not unreal, the microstructure generated by the Poisson-Voronoi tessellation is initialised in a multiphase-field framework~\cite{perumal2017phase,perumal2018phase}.

The entire approach employed in developing the topology-based grain detecting model is shown in Fig.\ref{fig:fig1}. 
Using Poisson-Voronoi tessellation, five \textit{populous} microstructures, each comprising of approximately $500$ grain with average size of $50\mu m$ are developed. 
Each of these microstructure, that constitute the principal dataset, are sliced to make them computationally \lq manageable\rq \thinspace.
In machine-learning based image analysis of microstructure, slicing the larger microstructures into smaller sections without the loss of any characteristic feature has generally been a common practise~\cite{holm2020overview}. 
Besides making the data manageable, the section of larger microstructures reduce the training efforts and seemingly increase the magnitude of data. 

In the current investigation, as shown in Fig.\ref{fig:fig1}, each larger microstructure is sliced into $64$ sections. 
This slicing ensures that atleast fifteen individual, and topologically-distinct, grains are included in each section.
Owing to the sectioning, the entire dataset comprised of $320$ smaller, yet polycrystalline, microstructure. 
While a significant number of the smaller microstructures ($80\%$) are involved in training the model, the validation is performed on the remnant ($20\%$).
Besides the numerically-established microstructures, polycrystalline images from open searches are also included to train the algorithm.

\begin{figure}
    \centering
      \begin{tabular}{@{}c@{}}
      \includegraphics[width=1.0\textwidth]{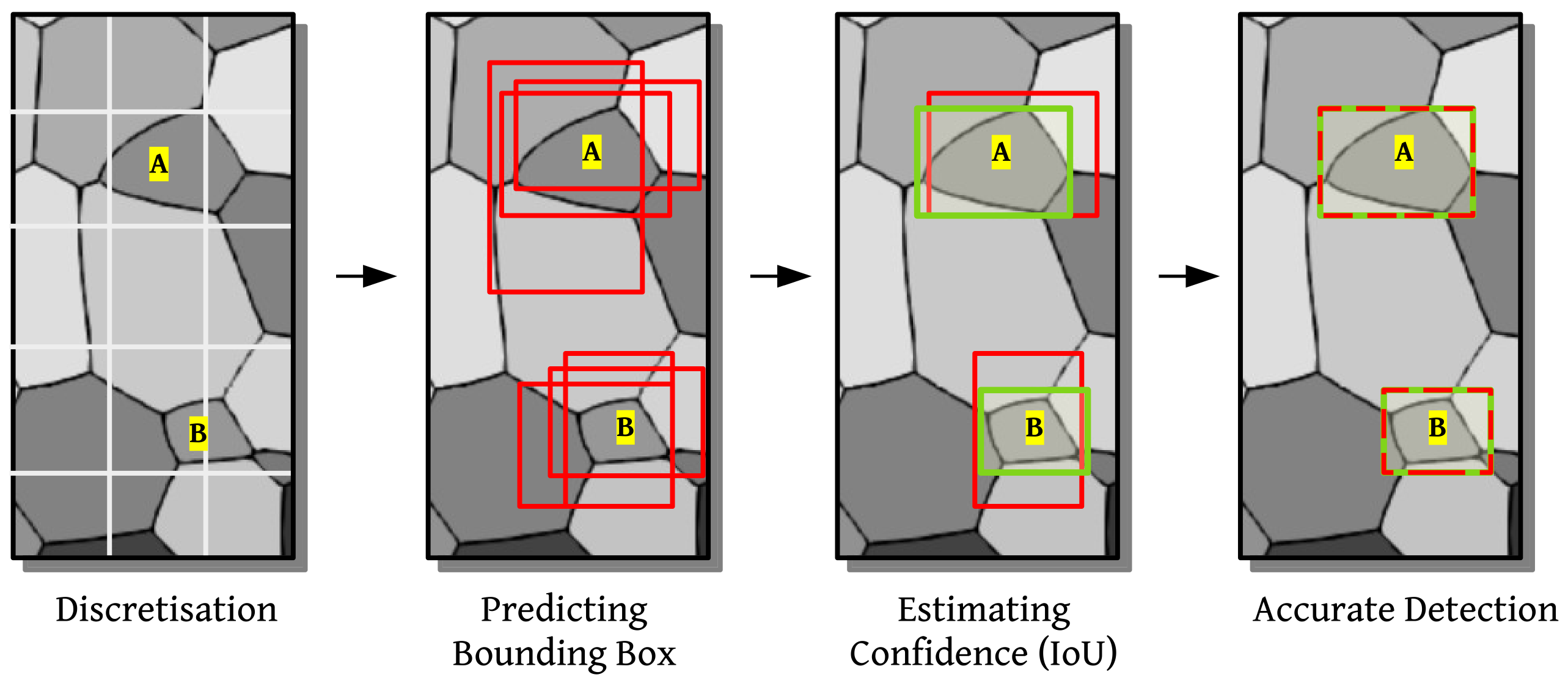}
    \end{tabular}
    \caption{  The approach adopted by the current object detection algorithm in realising grains of specific topological class in the polycrystalline continuum..
    \label{fig:fig2}}
\end{figure}

The present technique that detects the grains in a polycrystalline microstructure, while distinguishing their topological feature, is trained by manually locating the individual grains and labelling them based on the number of sides. 
A representative depiction of the manual demarcation of the grains, and the inclusion of the corresponding labels, is shown in Fig.\ref{fig:fig1}.
Using an appropriate tool, \textit{LabelImg}, the individual grains in the training dataset are located and labelled~\cite{tzutalin2015labelimg}. 
The spatial position of a given grain the microstructure is registered by introducing a suitable bounding box around it.
Moreover, the labels are appropriately titled to indicate the topological feature of the grains.
The bounding boxes along with the labels serve as the \textit{ground truth} of the microstructure while training the model.
Considering that, in two microstructures, number of sides assumed by a grain generally varies from four to seven, these face classes are exclusively focused in this analysis. 
Consequently, topological features of the grains constituting the microstructure are largely distinguished based on one of these four labels indicating face-class four to seven. 
Close to 250 sectioned microstructures, each warranting on an average of fifteen individual
bounding boxes, are manually labelled for training.

\subsection{Operating principle}

The approach adopted by the underlying algorithm (YOLOv5) in detecting the grains based on their topological feature is illustrated in Fig.\ref{fig:fig2}. 
Owing to its \textit{regression-based formulation}, the present technique essentially predicts the spatial location of a given class of grain, and gradually improving its accuracy in the subsequent iteration~\cite{jiang2022review}. 
This approach, by obviating the need for scanning the entire microstructure, ensures the optimal use of the computational resource. 

As indicated in Fig.\ref{fig:fig2}, the topology-based grain detection treatment begins with the discretisation of the microstructure into identically-sized $M\times M$ grids. 
Each of these grids serve as the spatial pivot for the predicting the position of the object initially. 
Following the discretisation of the microstructure, an assumption is made concerning the center of the object. 
In the present consideration, the center of the object is typically a point within the grain which is equidistant from all its faces. 
For a grain of specific face-class, assuming that its equidistant point coincides with the respective center of the grids, each discretised cell predicts a \textit{bounding box} with the intent to envelop the entire object.  
Consequently, as shown in Fig.\ref{fig:fig2}, the initial prediction yields several bounding boxes encircling a given object. 

By their predictions, each grid generates a vector, $\textbf{b}\in \{\text{b}_x, \text{b}_y, \text{b}_w, \text{b}_h \}$, which includes the spatial coordinates of the centre of the bounding box ($\text{b}_x$ and $ \text{b}_y$),  along with their dimensions as breath and height $\text{b}_x$ and $\text{b}_y$.
The components of the predicted vector characteristically vary with each discretised cell, despite all bounding boxes encompassing a specific grain (object). 
The subsequent steps following the introduction of the predictive bounding boxes involve identifying the most suitable encapsulation and refining its dimension to accurately capture the position of the object. 
To that end, the predictions of the discretised grids are analysed in view of the ground truth established through the manual labelling. 
The agreement between the prediction and the ground truth is quantified as \textit{confidence}~\cite{thomas2015high}.

Initially, the confidence is estimated for all predicted bounding boxes, irrespective of the class of grains they encompass, through the relation
\begin{equation}
\mathcal{C}_{\text{box}}=\mathcal{P}(\text{ob})\times IoU.
\end{equation}
The above expression yields \textit{box confidence} ($\mathcal{C}_{\text{box}}$) by combining the probability of locating any object, irrespective of its class, in a predicted box, $\mathcal{P}(\text{ob})$, with the Intersection over Union, $IoU$.
While the probability of encapsulating an object is dictated by vector components associated with the bounding box, the Intersection over Union quantifies the spatial overlap between the prediction and ground truth, as shown in Fig.\ref{fig:fig2}.
Given that the present model is developed to detect four different classes grains, a refined confidence of locating a desired class of object is ascertained. 
The resulting \textit{class confidence} reads
\begin{align}
 \mathcal{C}_{\text{class:}i} & =\mathcal{P}(\text{cl}_i|\text{ob}) \times \mathcal{C}_{\text{box}} \\ \nonumber 
 & = \mathcal{P}(\text{cl}_i|\text{ob}) \times \mathcal{P}(\text{ob})\times IoU,
\end{align}
with $\mathcal{P}(\text{cl}_i|\text{ob})$ indicating the condition probability of locating an object of class-$i$ within the bounding box. 
The estimated confidence for the four difference classes of grains are augmented to the vector $\textbf{b}$ which consequently reads
\begin{align}\label{bvector}
 \bm{b}=
 \begin{bmatrix}
    b_x\\
    b_y\\
    b_w\\
    b_h\\
    \mathcal{C}_{\text{FC4}}\\
    \vdots\\
    \mathcal{C}_{\text{FC7}}
\end{bmatrix}
\end{align}
where $\mathcal{C}_{\text{FC4}}$ and $\mathcal{C}_{\text{FC7}}$ correspondingly represent the class confidence of locating four- and seven-sided grains.

\begin{figure}
    \centering
      \begin{tabular}{@{}c@{}}
      \includegraphics[width=1.0\textwidth]{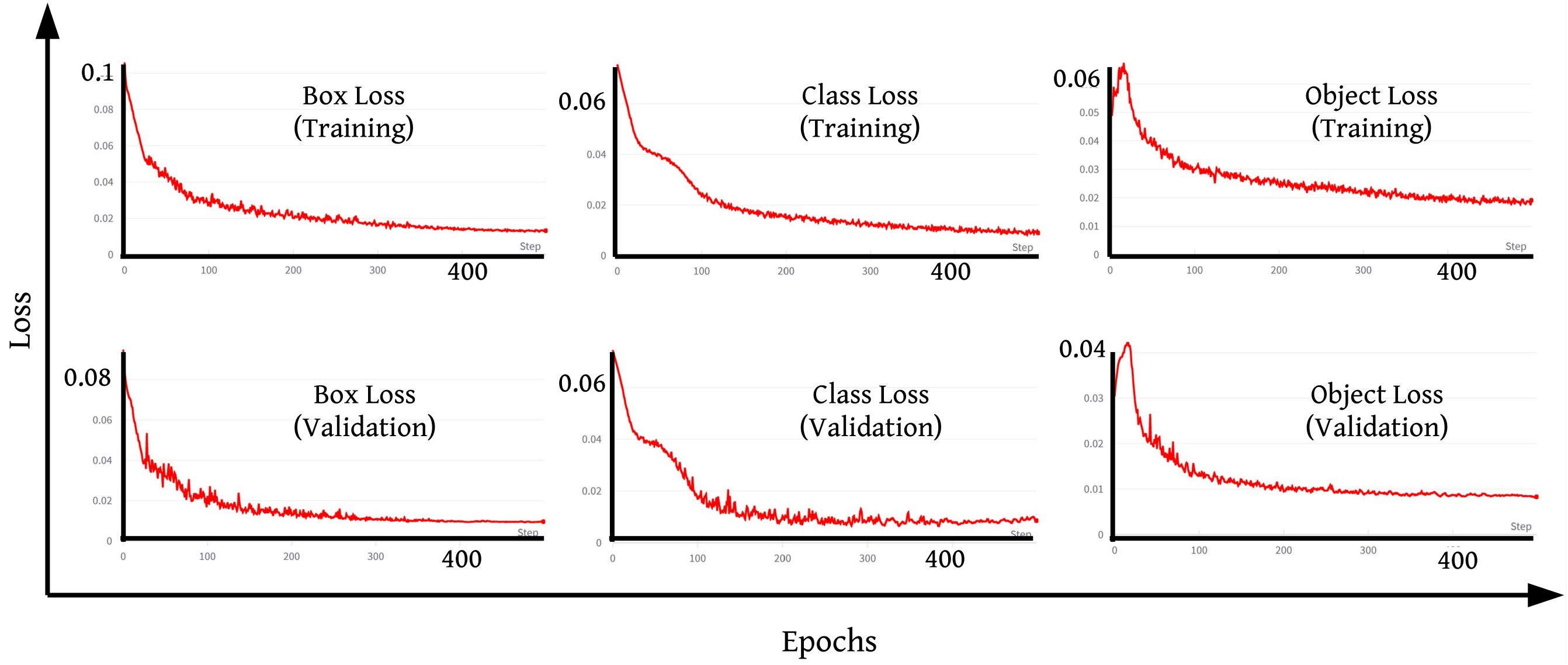}
    \end{tabular}
    \caption{ The deviation in the predicted topological class of the grain from the ground truth is quantified over the epochs using the loss functions. The decrease in the loss function for both test and validation dataset indicates the convergence of the model detection to the ground truth. 
    \label{fig:fig3}}
\end{figure}

The revised vector $\bm{b}$ is iteratively examined and varied in relation to the corresponding vector generated by the ground truth to perfect the bounding box detecting the grains of specific topological feature. 
As the vector $\bm{b}$ gets increasingly proximate to the ground truth, for a given bounding box, only the confidence associated with the specific class assumes a non-trivial value while other class confidences turn null.  
Principally based on the confidence, and involving \textit{non-max suppression} treatment, the best fitted bounding box is realised and refined~\cite{hosang2017learning}. 

\subsection{Comprehending accuracy}

Considering that grain boundaries in a polycrystalline system are inherently distinguishable from the bulk of the microstructure, a sophisticated visualisation tool could be adopted to isolate these interfaces.
Subsequently, suitable image analysis technique can be involved to examine the configuration of the segmented grain boundaries. 
Though such treatment is conceivable, quantifying its accuracy poses a challenge on its own. 
On the other hand, techniques built on a robust statistical framework facilitate, akin to the one employed in this investigation, offer definite ways of understanding the accuracy. 
To that end, the present detection approach and its corresponding outcomes are analysed through the associated methods in order to explicate the accuracy. 

\subsection{Minimising deviations}

Although the prediction of the algorithm is translated to a vector in Eqn.~\eqref{bvector}, this representation denotes bounding box generated by a discretised grid for an object of a specific class.
Correspondingly, the prediction vector assumes a multidimensional form depending on the number of discretised cells, objects, their classes and \textit{anchor boxes}. 
Despite the extensive dimensionality of the vector, the associated components can be compared with the ground truth to ascertain the accuracy of the prediction. 
Based primarily on the predicted bounding-box vector, and its corresponding components, the deviation from the ground truth can be quantified as \textit{loss} through suitable formulations called \textit{loss function}~\cite{huang2018yolo}.

The various forms of losses rendered by the present topological predictions are iteratively determined, over 500 epochs, for batches comprising of $15$ sliced-microstructures. 
The progressive change in the respective box, class and object loss across the epochs are graphically presented in Fig.~\ref{fig:fig3}. 
Monitoring the losses, particularly in the validation dataset, enables the fine tuning of the \textit{hyperparameters} and consequently, the accuracy of the treatment. 

Bounding box indicating the ground truth is compared with the prediction to estimate the box loss. 
Stated otherwise, the box loss quantifies the disparity between the ground truth and the predictions through the features of the bounding box including center, aspect ratio, dimensions and  Intersection over Union (IoU).
Object loss, on the other hand, focuses on the ability of the model to distinguish an object (a specific grain) from its background.  
Any deficiency in realising the grains in a microstructure is captured as the object loss. 
Inaccuracy in identifying the topological feature of the detected grains is expressed as the class loss. 
Both object and class loss are ascertained from the probabilities indicating the confidence in the bounding box vector through \textit{binary cross entropy}.
Irrespective of the nature of deviation the losses indicate, Fig.~\ref{fig:fig3} demonstrates that at the end of the 400 epochs the model is adequately trained, and exhibits marginal disparity from the ground truth. 
Furthermore, the minimal losses at the end of the epoch in the validation dataset assert the choice of the hyperparameters which include learning rate of 0.1 along with momentum and weight decay of 0.937 and 0.0005, respectively.

\subsection{Detection accuracy}

\begin{figure}
    \centering
      \begin{tabular}{@{}c@{}}
      \includegraphics[width=0.5\textwidth]{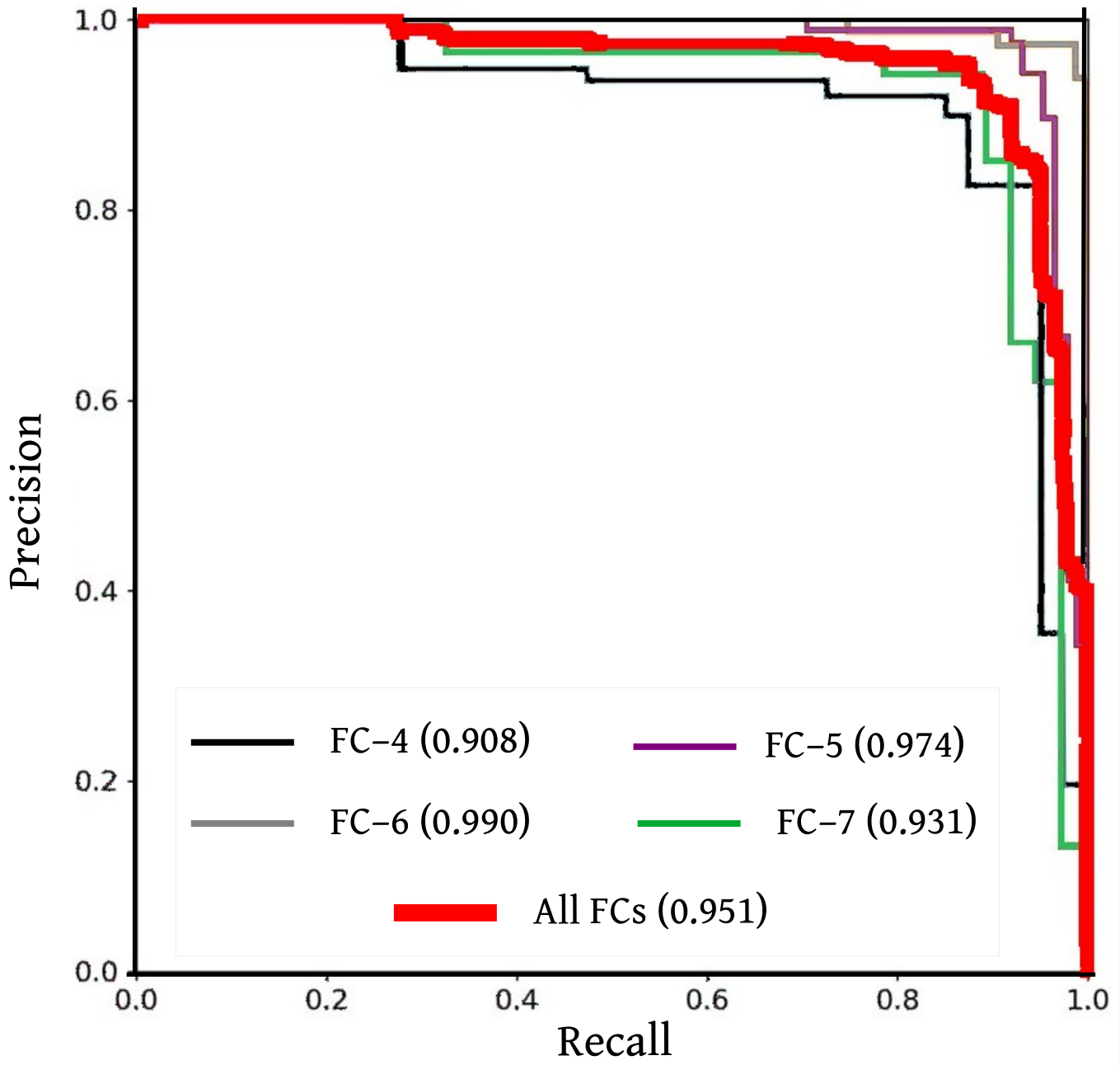}
    \end{tabular}
    \caption{ The rigour of the approach in making accurate topology-based detection of the grains is estimated as Precision and Recall. These parameters are collectively analysed for the individual face-classes, and the overall performance is ascertained from the mean average-precision (mAP) represented in dark-red line.
    \label{fig:fig4}}
\end{figure}

Considering that the model fundamentally detects grains along with their topological classes, besides losses, the accuracy of its performance are further investigated by closely examining the detections. 
The regression-based detections made by the current approach, in relation to the ground truth, can be categorised as true positive (TP), false positive (FP) and false negative (FN). 
While true positive is the accurate detection of the algorithm, the erroneous attribution to a face class different from the ground truth is referred as false positive. 
Similarly, false negative indicates failure to accurately detect a specific class of object. 
Using these categories of positives and negative, which extensively encompasses both accurate and faulty detections, the \textit{precision} of the model can be estimated by 
\begin{align}\label{Eqn:P}
 \text{P}=~\frac{\text{TP}}{\text{TP+FP}}.
\end{align}
Furthermore, an analogous parameter called \textit{recall} is determined through 
\begin{align}\label{Eqn:R}
 \text{R}=~\frac{\text{TP}}{\text{TP+FN}},
\end{align}
to gauge the performance of the detection treatment. 
As indicated by Eqn.~\eqref{Eqn:P}, precision realises the accuracy of the approach by augmenting inaccurate detection to the true positive in the formulation. 
Recall, on the other hand, describes the performance in the context of failure to definitively detect the existing face-class, \textit{id est} false negative. 

Both the parameters, precision and recall are collectively examined in relation to one other to demonstrate the accuracy of the present topology-detection model, as shown in Fig.~\ref{fig:fig4}~\cite{buckland1994relationship}. 
When the detection is absolute, with marginal inaccuracies in the form of false positive and negative, the corresponding precision-recall (PR) curve establishes an area of 1.0, encapsulating the entire plot.
Accordingly, any deviation from the complete encapsulation reflects the non-compliance of the detection with the ground truth, which can be ascertained from the area under the PR curve.
Considering that the terms contributing to the precision and recall (Eqns.~\eqref{Eqn:P} and ~\eqref{Eqn:R}) are realised for individual object (face) class, four distinct PR-curves constitute the graphical representation in Fig.~\ref{fig:fig4}.
It is evident from this depiction that the present algorithm renders seemingly convincing detection of grains with topological classes ranging from four to seven. 
Owing to its increased presence in the polycrystalline microstructure, the model detects six-sided grains more accurately when compared to the rest of the face-classes. 
Furthermore, the solid red line in Fig.~\ref{fig:fig4} indicates average of the PR-curves pertaining to the individual face-classes.
The overall performance of the model is  affirmed by the red mean average-precision (mAP) curve through its predominant encapsulation ($95\%$) of the PR plot~\cite{torgo2009precision}. 

\begin{figure}
    \centering
      \begin{tabular}{@{}c@{}}
      \includegraphics[width=0.5\textwidth]{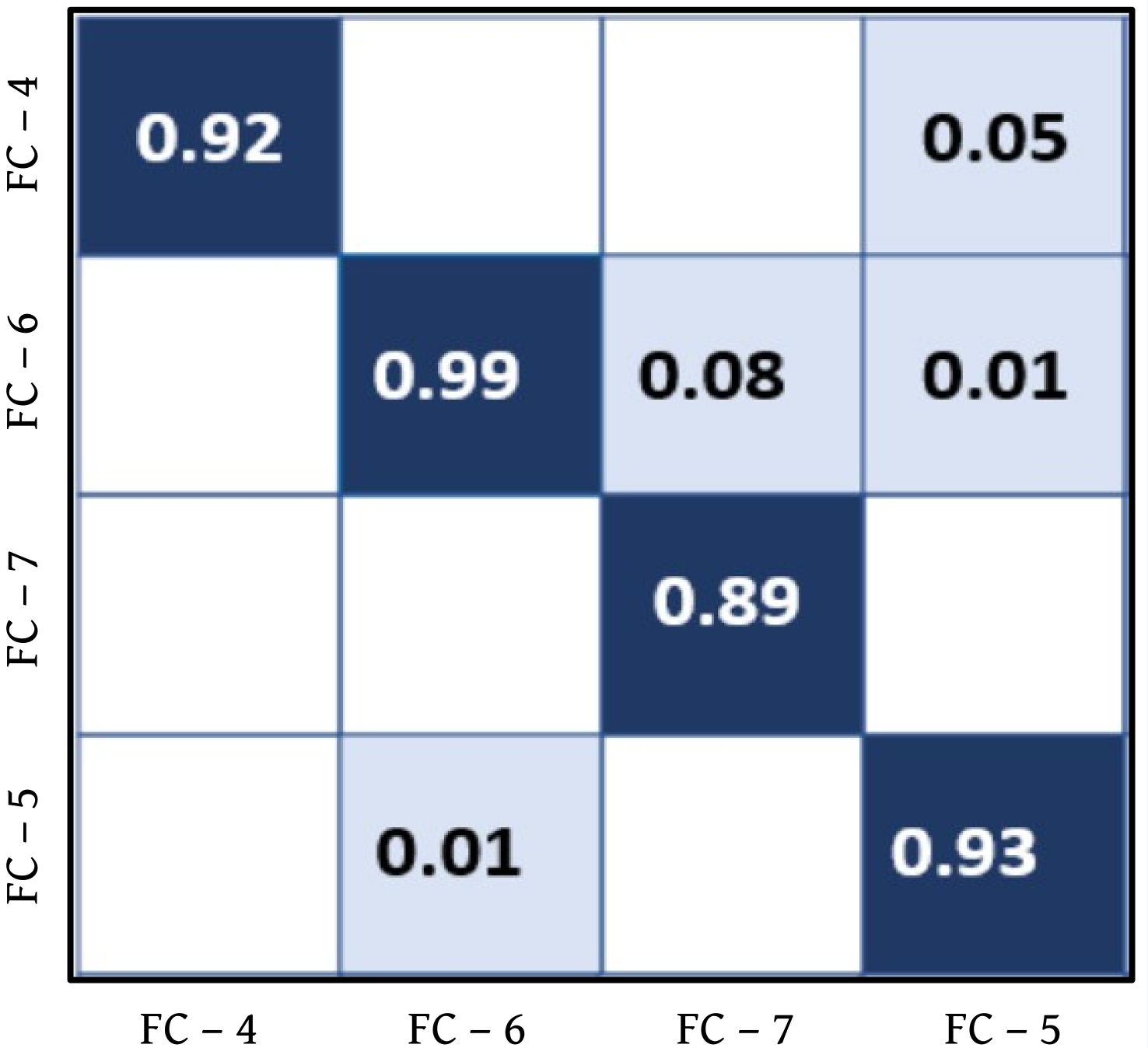}
    \end{tabular}
    \caption{ Besides the accurate detections represented by the diagonal entries, the inaccurate predictions in the form of false-positives and -negatives, across various topological classes, are \textit{normalised} and presented as a confusion matrix. In the present representation, a blank entry indicates insignificant contribution to the overall detections.
    \label{fig:fig5}}
\end{figure}

Before exposing the trained and validated model to detect grains in unknown microstructure, model is examined for any \textit{biases}~\cite{salmon2015proper}. 
In other words, when employed to detect object of different classes in a microstructure, the underlying algorithm can get skewed towards (or against) a specific class.
In order to ensure that such biases are rather absent in the present, a relevant confusion matrix is devised and shown in Fig.~\ref{fig:fig5}. 
The diagonal elements of the confusion matrix expresses the true positive, the components of false positive and false negative across the different face classes are expressed as the non-diagonal entries. 
The confusion matrix rendered by the present model in Fig.~\ref{fig:fig5} unravels that the model is not biased towards any particular face-class and the true positive significantly dominates inaccuracies of any form. 

\section{Results and Discussion}

The trained and validated algorithm is employed to detect grains in the polycrystalline microstructures, which have not been involved in the development of the approach thus far.
In order to ensure that the performance of the model is not restricted to the number density of the objects, the testing is extended to both sliced microstructure with dimensions identical to the training data, and larger polycrystalline systems with significantly greater number of grains. 

\begin{figure}
\centering
   \begin{subfigure}[b]{1.0\textwidth}
   \includegraphics[width=1\linewidth]{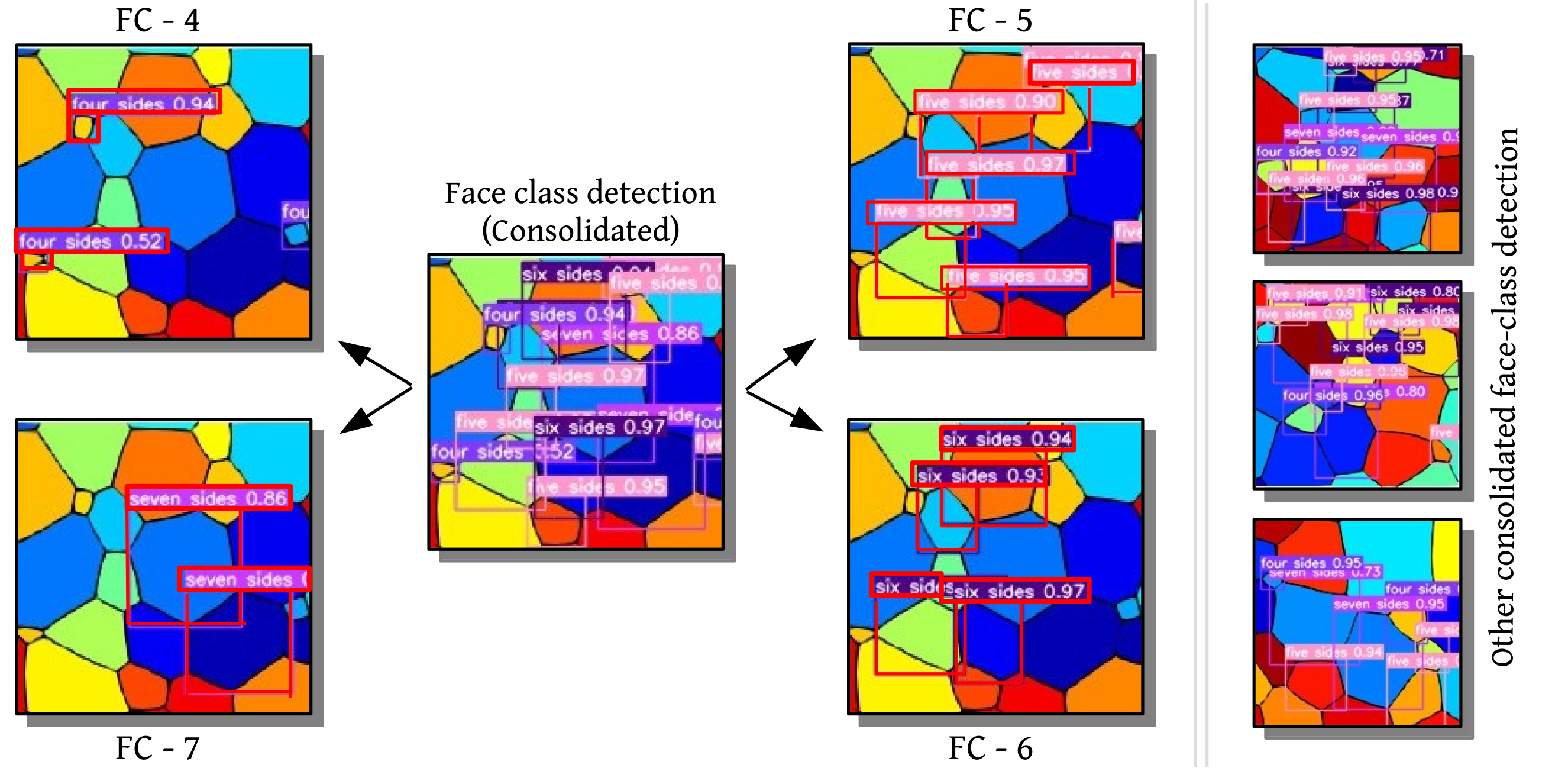}
   \caption{}
   \label{fig:fig6a} 
\end{subfigure}

\begin{subfigure}[b]{1.0\textwidth}
   \includegraphics[width=1\linewidth]{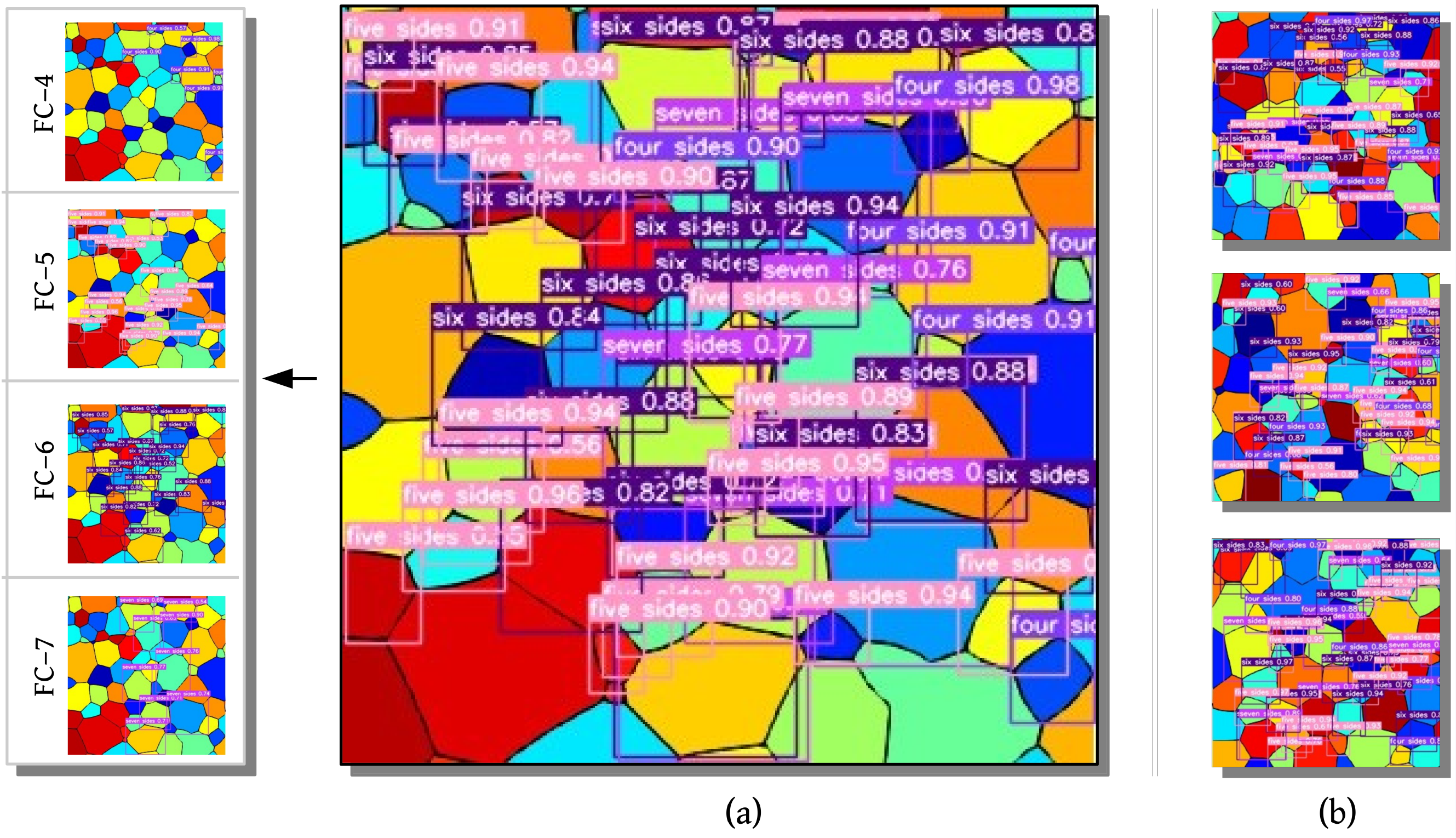}
   \caption{}
   \label{fig:fig6b}
\end{subfigure}

\caption[Two numerical solutions]{ (a) Topology-based detection rendered by the model in a sliced microstructure with fewer grains. The realisation of grains with specific number of sides are presented separately. (b) Adopting the trained model to analyse the topological features of  larger microstructures with increased number of grains. }
\end{figure}

\subsection{Detection face-classes}

In Fig.~\ref{fig:fig6a}, the performance of the model in detecting grains based on their topological feature across the smaller sliced microstructure is illustrated. 
In this representative depiction of the four sliced polycrystalline systems, detection of grains in one of the microstructure is expanded and presented individually based on their topological class. 
The relatively low number-density of grains constituting the microstructures in Fig.~\ref{fig:fig6a} lends itself to a rather direct assessment of the model performance. 
Stated otherwise, besides the various statistical metrics, the accuracy of the model in making topology-based detection across the previously unknown sliced microstructures can be verified against the unaided conventional perception. 
Such cross-examination of detections in Fig.~\ref{fig:fig6a} explicates the convincing performance of the model.

\subsection{Applicability to larger systems}

The test data for analysing the detection approach is replaced with microstructures of larger dimension encompassing increased number of grains. 
The topology-based grain detection rendered by the algorithm in representative large-polycrystalline systems is shown in Fig.~\ref{fig:fig6b}. 
Considering that each of these microstructures include atleast 50 detectable grains, a direct cross-examination of  the performance is rather an arduous task. 
However, from the density of the bounding boxes in Fig.~\ref{fig:fig6b}, it is indicative that the model captures almost all the grains, irrespective of the size of the microstructure and number of grains it accommodates. 
The seemingly unwavered performance of the model, despite the increased population of grains, affirms the versatile applicability of the model.

\subsection{EBSD microstructures}

The training, validation and testing of the present topology-based grain detection model has predominantly dictated by numerically-devised polycrystalline microstructure. 
Owing to underlying scheme, though the simulated microstructures can be claimed to principally reflect the experimentally observed polycrystalline microstructure, the definitive performance of the model can only gauged when it is exposed to physical systems. 
To that end, besides analysing the topological features of the grains in polycrystalline systems rendered by \textit{open searches}, which is collectively presented in the supplementary document, a set of rather complex microstructures are investigated using the present approach. 

\begin{figure}
    \centering
      \begin{tabular}{@{}c@{}}
      \includegraphics[width=1.0\textwidth]{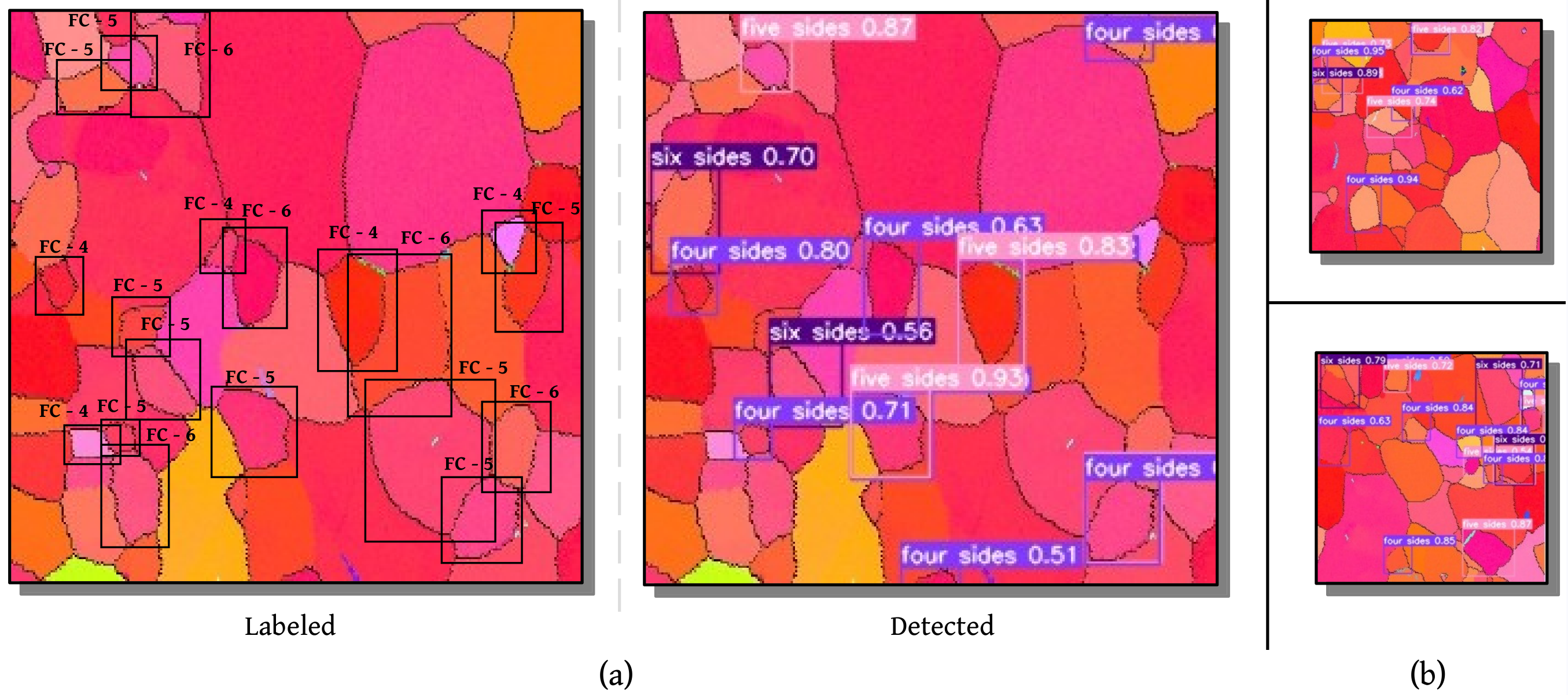}
    \end{tabular}
    \caption{  a) Manually labeled and detected topological features of a textured polycrystalline microstructure. b) Detection rendered by the present technique on other textured micrographs.
    \label{fig:fig7}}
\end{figure}

Desired properties in a material are generally established by exposing to suitable processing technique. 
These processing techniques achieve  the desired behaviour by characteristically altering the microstructure. 
In polycrystalline materials, the configuration of the constituent grains get significantly changed to the extent that these microstructures exhibit noticeable deviation from the numerically-generated ones. 
Particularly, when the polycrystalline materials are exposed to large-scale deformation processes, like rolling, the conventional grain boundary structures get distorted and a typifying texture is introduced.
Stated otherwise, as opposed to generally straight faces connecting triple junctions, curved grain boundaries are not uncommon in rolled materials. 
The performance of the present model in handling textured microstructures is analysed by considering Electron Back Scattered Diffraction (EBSD) micrographs of rolled magnesium alloy (AZ31) sample.
For direct comparison with the ground truth, in Fig.~\ref {fig:fig7} the topology based grain detection of the model is placed against the manual-labelled textured microstructure of AZ31 sample.
Evidently, the model captures major portion of the grains along with its topological feature, despite the characteristic change in the morphology of the grain and the variation in the colour scheme generally associated with the numerically-generated polycrystalline systems.
Besides, this investigation further unravels that performance of the model is rather inadequate when the boundaries of the grains are curved and ill-defined. 
The model offers similar outcomes on other textured EBSD micrographs included in  Fig.~\ref {fig:fig7}. 
This rather atypical performance of the present technique on textured microstructures can be attributed to the nature of the training and validation dataset which primarily comprised of pristine numerically-generated microstructure. 
Based on the performance of the algorithm in the conventional microstructure, it can reasonably be stated that the accuracy of detection in deformed systems can be improved by adopting sufficient number of the textured microstructures in the development.

\section{Conclusion}

Image processing techniques have long since been employed to quantitatively describe a microstructure.
With of advent artificial intelligence and related algorithm, aide offered by such external treatments have increased considerably.
Though recently the sophisticated techniques of deep learning are progressively employed in microstructure analysis, their role in extracting features of polycrystalline systems is rather marginal. 
Therefore, in the present work, a regression-based object counting algorithm is trained, and developed, for comprehending the topological features of the grains constituting polycrystalline microstructures. 

The topological features of a given polycrystalline system, in the current approach, is ascertained by detecting and counting grains based on their face classes. 
In other words, the algorithm is trained to examine individual grains of the polycrystalline microstructure, and handle them as distinct \textit{object-classes} depending on the number of sides. 
Correspondingly, the grains with five sides are detected and counted as a specific object class, as opposed face-class six or seven grains. 
When the resulting approach is allowed to explore the entire microstructure, it yields the spatial and numerical distribution of the grains based on the topological feature. 
While the spatial distribution is rendered by the detection aspect of the algorithm, the object counting offers the number density of individual face-class in a polycrystalline system.

Though the present approach, owing to its regression based algorithm, is efficient in detecting the topological features of the polycrystalline system, its accuracy relies heavily on the training and validation dataset. 
Correspondingly, despite performing convincingly in polycrystalline microstructures with definite and flat grain boundaries, the deviation from the ground truth becomes noticeable while handling processed grains with curved interfaces. 
This limitation of the detection treatment will be explored in the upcoming works by primarily examining microstructure of systems exposed to various processing techniques.

\section*{Data Availability}

As a key feature, the model and dataset developed in the present investigation is made available for its usage, particularly on experimentally-observed microstructures, in:\\
\url{https://github.com/theormets/Topological-features-of-grains}.

\section*{Acknowledgments}

PGK Amos thanks the financial support of the SCIENCE \& ENGINEERING RESEARCH BOARD (SERB) under the project SRG/2021/000092.

\bibliographystyle{elsarticle-num}
\bibliography{library.bib}

\begin{thebibliography}{10}

\bibitem{deschamps2021precipitation}
Alexis Deschamps and CR~Hutchinson.
\newblock Precipitation kinetics in metallic alloys: Experiments and modeling.
\newblock {\em Acta Materialia}, 220:117338, 2021.

\bibitem{rohrer2005influence}
Gregory~S Rohrer.
\newblock Influence of interface anisotropy on grain growth and coarsening.
\newblock {\em Annual Review of Materials Research}, 35(1):99--126, 2005.

\bibitem{armstrong1970influence}
RW~Armstrong.
\newblock {\em Metallurgical and Materials Transactions B}, 1(5):1169--1176,
  1970.

\bibitem{kim2006computer}
Seong~Gyoon Kim, Dong~Ik Kim, Won~Tae Kim, and Yong~Bum Park.
\newblock Computer simulations of two-dimensional and three-dimensional ideal
  grain growth.
\newblock {\em Physical Review E}, 74(6):061605, 2006.

\bibitem{perumal2020quadrijunctions}
Ramanathan Perumal, PG~Kubendran Amos, Michael Selzer, and Britta Nestler.
\newblock Quadrijunctions-stunted grain growth in duplex microstructure: a
  multiphase-field analysis.
\newblock {\em Scripta Materialia}, 182:16--20, 2020.

\bibitem{amos2020multiphase}
PG~Kubendran Amos, Ramanathan Perumal, Michael Selzer, and Britta Nestler.
\newblock Multiphase-field modelling of concurrent grain growth and coarsening
  in complex multicomponent systems.
\newblock {\em Journal of Materials Science \& Technology}, 45:215--229, 2020.

\bibitem{mullins1956two}
William~W Mullins.
\newblock Two-dimensional motion of idealized grain boundaries.
\newblock {\em Journal of Applied Physics}, 27(8):900--904, 1956.

\bibitem{glazier1993grain}
James~A Glazier.
\newblock Grain growth in three dimensions depends on grain topology.
\newblock {\em Physical Review Letters}, 70(14):2170, 1993.

\bibitem{von1952metal}
John Von~Neumann.
\newblock Metal interfaces.
\newblock {\em American Society for Metals, Cleveland}, 108, 1952.

\bibitem{macpherson2007neumann}
Robert~D MacPherson and David~J Srolovitz.
\newblock The von neumann relation generalized to coarsening of
  three-dimensional microstructures.
\newblock {\em Nature}, 446(7139):1053--1055, 2007.

\bibitem{patterson2022relationship}
Burton~R Patterson.
\newblock Relationship between mean grain face curvature and number of faces in
  normal grain growth: The meaning of normalized integral mean curvature.
\newblock {\em Acta Materialia}, 229:117724, 2022.

\bibitem{wakai2000three}
Fumihiro Wakai, Naoya Enomoto, and Hiroshi Ogawa.
\newblock Three-dimensional microstructural evolution in ideal grain
  growth—general statistics.
\newblock {\em Acta Materialia}, 48(6):1297--1311, 2000.

\bibitem{rivier1985statistical}
N~Rivier.
\newblock Statistical crystallography structure of random cellular networks.
\newblock {\em Philosophical Magazine B}, 52(3):795--819, 1985.

\bibitem{ullah2017simulations}
Asad Ullah, Matiullah Khan, Xue Weihua, Safdar Hussain, Mujeeb ur~Rahman,
  Nadeem Salamat, Fazal Haq, et~al.
\newblock Simulations of grain growth in realistic 3d polycrystalline
  microstructures and the macpherson--srolovitz equation.
\newblock {\em Materials Research Express}, 4(6):066502, 2017.

\bibitem{alkemper2001quantitative}
J~Alkemper and PW1827210 Voorhees.
\newblock Quantitative serial sectioning analysis.
\newblock {\em Journal of microscopy}, 201(3):388--394, 2001.

\bibitem{groeber20063d}
Michael~A Groeber, BK~Haley, Michael~D Uchic, Dennis~M Dimiduk, and Somnath
  Ghosh.
\newblock 3d reconstruction and characterization of polycrystalline
  microstructures using a fib--sem system.
\newblock {\em Materials characterization}, 57(4-5):259--273, 2006.

\bibitem{bhandari20073d}
Y~Bhandari, S~Sarkar, M~Groeber, MD~Uchic, DM~Dimiduk, and S~Ghosh.
\newblock 3d polycrystalline microstructure reconstruction from fib generated
  serial sections for fe analysis.
\newblock {\em Computational Materials Science}, 41(2):222--235, 2007.

\bibitem{ullah2013optimal}
Asad Ullah, Guoquan Liu, Hao Wang, Matiullah Khan, Dil~Faraz Khan, and Junhua
  Luan.
\newblock Optimal approach of three-dimensional microstructure reconstructions
  and visualizations.
\newblock {\em Materials Express}, 3(2):109--118, 2013.

\bibitem{sha2020artificial}
Wuxin Sha, Yaqing Guo, Qing Yuan, Shun Tang, Xinfang Zhang, Songfeng Lu, Xin
  Guo, Yuan-Cheng Cao, and Shijie Cheng.
\newblock Artificial intelligence to power the future of materials science and
  engineering.
\newblock {\em Advanced Intelligent Systems}, 2(4):1900143, 2020.

\bibitem{amos2022high}
PG~Kubendran Amos, Arnd Koeppe, Ramanathan Perumal, and Britta Nestler.
\newblock High-fidelity simulations and data-driven insights on rate-governing
  phases in duplex and triplex systems during isotropic normal grain growth.
\newblock {\em Physical Review Materials}, 6(11):113401, 2022.

\bibitem{decost2015computer}
Brian~L DeCost and Elizabeth~A Holm.
\newblock A computer vision approach for automated analysis and classification
  of microstructural image data.
\newblock {\em Computational materials science}, 110:126--133, 2015.

\bibitem{azimi2018advanced}
Seyed~Majid Azimi, Dominik Britz, Michael Engstler, Mario Fritz, and Frank
  M{\"u}cklich.
\newblock Advanced steel microstructural classification by deep learning
  methods.
\newblock {\em Scientific reports}, 8(1):1--14, 2018.

\bibitem{zou2019object}
Zhengxia Zou, Zhenwei Shi, Yuhong Guo, and Jieping Ye.
\newblock Object detection in 20 years: A survey.
\newblock {\em arXiv preprint arXiv:1905.05055}, 2019.

\bibitem{thuan2021evolution}
Do~Thuan.
\newblock Evolution of yolo algorithm and yolov5: The state-of-the-art object
  detention algorithm.
\newblock 2021.

\bibitem{kumar1992properties}
Susmit Kumar, Stewart~K Kurtz, Jayanth~R Banavar, and MG~Sharma.
\newblock Properties of a three-dimensional poisson-voronoi tesselation: A
  monte carlo study.
\newblock {\em Journal of statistical physics}, 67(3):523--551, 1992.

\bibitem{perumal2017phase}
Ramanathan Perumal, PG~Kubendran Amos, Michael Selzer, and Britta Nestler.
\newblock Phase-field study on the formation of first-neighbour topological
  clusters during the isotropic grain growth.
\newblock {\em Computational Materials Science}, 140:209--223, 2017.

\bibitem{perumal2018phase}
Ramanathan Perumal, PG~Kubendran Amos, Michael Selzer, and Britta Nestler.
\newblock Phase-field study of the transient phenomena induced by
  ‘abnormally’large grains during 2-dimensional isotropic grain growth.
\newblock {\em Computational Materials Science}, 147:227--237, 2018.

\bibitem{holm2020overview}
Elizabeth~A Holm, Ryan Cohn, Nan Gao, Andrew~R Kitahara, Thomas~P Matson,
  Bo~Lei, and Srujana~Rao Yarasi.
\newblock Overview: Computer vision and machine learning for microstructural
  characterization and analysis.
\newblock {\em Metallurgical and Materials Transactions A}, 51(12):5985--5999,
  2020.

\bibitem{tzutalin2015labelimg}
Tzutalin.
\newblock Labelimg.
\newblock Free Software: MIT License, 2015.

\bibitem{jiang2022review}
Peiyuan Jiang, Daji Ergu, Fangyao Liu, Ying Cai, and Bo~Ma.
\newblock A review of yolo algorithm developments.
\newblock {\em Procedia Computer Science}, 199:1066--1073, 2022.

\bibitem{thomas2015high}
Philip Thomas, Georgios Theocharous, and Mohammad Ghavamzadeh.
\newblock High confidence policy improvement.
\newblock In {\em International Conference on Machine Learning}, pages
  2380--2388. PMLR, 2015.

\bibitem{hosang2017learning}
Jan Hosang, Rodrigo Benenson, and Bernt Schiele.
\newblock Learning non-maximum suppression.
\newblock In {\em Proceedings of the IEEE conference on computer vision and
  pattern recognition}, pages 4507--4515, 2017.

\bibitem{huang2018yolo}
Rachel Huang, Jonathan Pedoeem, and Cuixian Chen.
\newblock Yolo-lite: a real-time object detection algorithm optimized for
  non-gpu computers.
\newblock In {\em 2018 IEEE International Conference on Big Data (Big Data)},
  pages 2503--2510. IEEE, 2018.

\bibitem{buckland1994relationship}
Michael Buckland and Fredric Gey.
\newblock The relationship between recall and precision.
\newblock {\em Journal of the American society for information science},
  45(1):12--19, 1994.

\bibitem{torgo2009precision}
Luis Torgo and Rita Ribeiro.
\newblock Precision and recall for regression.
\newblock In {\em International Conference on Discovery Science}, pages
  332--346. Springer, 2009.

\bibitem{salmon2015proper}
Brian~P Salmon, Waldo Kleynhans, Colin~P Schwegmann, and Jan~C Olivier.
\newblock Proper comparison among methods using a confusion matrix.
\newblock In {\em 2015 IEEE International Geoscience and Remote Sensing
  Symposium (IGARSS)}, pages 3057--3060. IEEE, 2015.

\end{thebibliography}
\end{document}